\input harvmac
\input epsf


\lref\rWitten{A. Hanany and E. Witten, {\it Type IIB Superstrings, BPS
Monopoles and three dimensional gauge dynamics,} Nucl. Phys. {\bf B492}
(1997) 152-190, {\tt hep-th/9611230}
\semi
E. Witten, {\it Solutions of four-dimensional field theories
via M-theory,} Nucl. Phys. {\bf B500} (1997) 3-42, {\tt hep-th/9703166}.}
\lref\rGukov{S. Gukov, {\it Seiberg-Witten Solution from Matrix Theory,}
{\tt hep-th/9709138}
\semi
A. Kapustin, {\it Solution of N=2 Gauge Theories via Compactification to
three dimensions,} Nucl. Phys. {\bf B534} (1998) 531-545, 
{\tt hep-th/9804069}
\semi
R. de Mello Koch and J.P. Rodrigues, {\it Solving four dimensional field
theories with the Dirichlet fivebrane,} Phys.Rev. {\bf D60} (1999) 027901, 
{\tt hep-th/9811036}.}
\lref\rOz{N.D. Lambert and P.C. West, {\it N=2 Superfields and the 
M-Fivebrane,} Phys. Lett. {\bf B424} (1998) 281, {\tt hep-th/9801104}
\semi
N.D. Lambert and P.C. West, {\it Gauge Fields and M-Fivebrane Dynamics,}
Nucl. Phys. {\bf B524} (1998) 141, {\tt hep-th/9712040},
\semi
P.S. Howe, N.D. Lambert and P.C. West, {\it Classical M-Fivebrane Dynamics
and Quantum N=2 Yang-Mills,} Phys. Lett. {\bf B418} (1998) 85,
{\tt 9710034}
\semi
J. de Boer, K. Hori, H. Ooguri and Y. Oz, {\it Kahler Potential and Higher
Derivative terms from M Theory Fivebrane,} Nucl. Phys. {\bf B518} (1998)
173, {\tt hep-th/9711143}.}
\lref\rOoguri{E. Witten, {\it Branes and the Dynamics of QCD,} Nucl. Phys.
{\bf B507} (1997) 658, {\tt hep-th/9706109}
\semi
H. Ooguri, {\it M Theory Fivebrane and SQCD,} Nucl.Phys.Proc.Suppl. 68 (1998)
84, {\tt hep-th/9709211}.}
\lref\rOfer{O. Aharony, A. Fayyazuddin and J. Maldacena, {\it The Large N
Limit of N=2,1 Field Theories from Threebranes in F-theory,} JHEP 9807
(1998) 013, {\tt hep-th/9806159}.}
\lref\rRadu{R. de Mello Koch and R. Tatar, {\it Higher Derivative Terms
from Threebranes in F Theory,} Phys.Lett. {\bf B450} (1999) 99, 
{\tt hep-th/9811128}.}
\lref\rUs{R. de Mello Koch, A. Paulin-Campbell and J.P. Rodrigues,
{\it Non-holomorphic Corrections from Threebranes in F Theory,} to appear in Phys.Rev.D,
{\tt hep-th/9903029}.}
\lref\rSen{A. Sen, {\it BPS States on a Threebrane Probe,} 
{\tt hep-th/9608005}.}
\lref\rFayy{A. Fayyazuddin, {\it Results in supersymmetric field theory
from 3-brane probe in F-theory,} Nucl. Phys. {\bf B497} (1997) 101,
{\tt hep-th/9701185}.}
\lref\rES{E. Witten and L. Susskind, {\it The Holographic Bound in Anti-de 
Sitter Space,} {\tt hep-th/9805114.}}
\lref\rJe{A. Jevicki and T. Yoneya, {\it Spacetime Uncertainty Principle
and Conformal Symmetry in D-particle Dynamics,} Nucl. Phys. {\bf B535}
(1998) 335, {\tt hep-th/9805069}
\semi
A. Jevicki, Y. Kazama and T. Yoneya, {\it Quantum Metamorphosis
of Conformal Transformation in D3 Brane Yang-Mills Theory,}
Phys. Rev. Lett. {\bf 81} (1998) 5072, {\tt hep-th/9808039}
\semi
A. Jevicki, Y. Kazama and T. Yoneya, {\it Generalized Conformal Symmetry
in D Brane Matrix Models,} Phys. Rev. {\bf D59} (1999) 066001,
{\tt hep-th/9810146}
\semi
F. Gonzalez-Rey, B. Kulik, I.Y. Park and M. Rocek, {\it Selfdual 
Effective Action of N=4 super Yang-Mills,} {\tt hep-th/9810152}.}
\lref\rSW{N. Seiberg and E. Witten, {\it Electric-Magnetic Duality, Monopole
Condensation and Confinement in N=2 Super Yang-Mills Theory,} 
Nucl. Phys. {\bf B426} (1994) 19, {\tt hep-th/9407087}
\semi
N. Seiberg and E. Witten, {\it Monopoles, Duality and Chiral
Symmetry Breaking in N=2 Supersymmetric QCD,} Nucl. Phys. {\bf B431}
(1994) 484, {\tt hep-th/9408099.}}
\lref\rFerrari{F. Ferrari, {\it The dyon spectra of finite gauge theories,}
Nucl. Phys. {\bf B501} (1997) 53, {\tt hep-th/9608005}.}
\lref\rLerche{W. Lerche, {\it Introduction to Seiberg-Witten Theory and its
Stringy Origin,} Nucl. Phys. Proc. Suppl. {\bf 55B} (1997) 83,
{\tt hep-th/9611190.}}
\lref\rWit{E. Witten, {\it New Perspectives in the Quest for Unification,}
{\tt hep-th 9812208} 
\semi
I. Klebanov, {\it From Threebranes to Large N Gauge Theories,}
{\tt hep-th 9901018}.}
\lref\rGubser{A. Kehagias and K. Sfetsos, {\it On Running Couplings in Gauge
Theories from IIB supergravity,} {\tt hep-th/9902125}
\semi
S.S. Gubser, {\it Dilaton-driven Confinement,} {\tt hep-th/9902155}
\semi
H. Liu and A.A.Tseytlin, {\it D3-brane-D-Instanton configuration and
N=4 Super YM Theory in constant self-dual background,} 
{\tt hep-th/9903091}
\semi
A. Kehagias and K. Sfetsos, {\it On Asymptotic Freedom and Confinement
from Type IIB Supergravity,} {\tt hep-th/9903109}
\semi
S. Nojiri and S.D. Odinstov, {\it Running gauge coupling and quark-antiquark
from dilatonic gravity,} {\tt hep-th/9904036}
\semi
S. Nojiri and S.D. Odinstov, {\it Two Boundaries AdS/CFT correspondence in 
dilatonic gravity,} Phys.Lett. {\bf B449} (1999) 39 {\tt hep-th/9904036}.}
\lref\rGaber{M.R. Gaberdiel, T. Hauer and B. Zweibach, {\it Open String-String junction transitions,}
Nucl. Phys. {\bf B525} (1998) 117, {\tt hep-th/9801205}.}
\lref\rBerg{O. Bergmann and A. Fayyazuddin, {\it String junctions and BPS states in Seiberg-Witten theory,}
Nucl.Phys. {\bf B531} (1998) 108, {\tt hep-th/9802033}
\semi
A. Mikhailov, N. Nekrasov and S.Sethi, {\it Geometric realisations of BPS states in N=2 theories,} Nucl.Phys. {\bf B531} (1998) 345,
{\tt hep-th/9803142}
\semi
O. De Wolfe and B. Zweibach, {\it String junctions for arbitrary Lie algebra representations,} Nucl. Phys. {\bf B541} (1999) 509,
{\tt hep-th/9804210}.}
\lref\rMal{C.G.Callan and J.M.Maldacena, {\it Brane Dynamics from the
Born-Infeld Action,} Nucl.Phys. {\bf B513} (1998) 198, 
{\tt hep-th/9708147}
\semi
G.W.Gibbons, {\it Born-Infeld Particles and Dirichlet p-branes,}
Nucl.Phys. {\bf B514} (1998) 603, {\tt hep-th/9709027}.}
\lref\rHash{A.Hashimoto, {\it Shape of Branes pulled by strings,}
Phys.Rev. {\bf D57} (1998) 6441, {\tt hep-th/9711097}.}
\lref\rChalm{G Chalmers, M. Rocek, and R. von Unge, {\it Monopoles in quantum corrected N=2 super Yang Mills theory},
{\tt hep-th/9612195}.}
\lref\rDenef{F. Denef, {\it Attractors at weak gravity,} {\tt hep-th/9812049}.}
\lref\rLowe{A. Sen, {\it F Theory and Orientifolds,} Nucl. Phys.
{\bf B475} (1996) 562, {\tt hep-th/9605150}
\semi
T.Banks, M.R.Douglas and N.Seiberg, {\it Probing F Theory with branes,}
Phys. Lett. {\bf B387} (1996) 74, {\tt hep-th/9605199}
\semi
M.R.Douglas, D.A.Lowe and J.H.Schwarz, {\it Probing F Theory with multiple
branes,} Phys. Lett. {\bf B394} (1997) 297, {\tt hep-th/9612062}.}
\lref\rSel{A.Fayyazuddin, {\it Some Comments on N=2 Supersymmetric Yang-Mills,}
Mod. Phys. Lett. A10 (1995) 2703, {\tt hep-th/9504120}
\semi
S.Sethi, M.Stern and E.Zaslow, {\it Monopole and Dyon Bound States in N=2
Supersymmetric Yang-Mills Theories,} Nucl. Phys. {\bf B457} (1995) 484,
{\tt hep-th/9508117}
\semi
J.P. Gauntlett and J.A. Harvey, {\it S-Duality and Dyon Spectrum in N=2 Super
Yang-Mills Theory,} Nucl.Phys. {\bf B463} (1996) 287, {\tt hep-th 9508156}
\semi
F.Ferrari and A.Bilal, {\it The Strong-Coupling Spectrum of the Seiberg-Witten
Theory,} Nucl. Phys. {\bf B469} (1996) 387, {\tt hep-th/9602082}
\semi
A.Klemm, W.Lerche, P.Mayr, C.Vafa and N.Warner, {\it Self-dual
Strings and N=2 Supersymmetric Field Theory,} Nucl. Phys. {\bf B477}
(1996) 746, {\tt hep-th/9604034}
\semi
F.Ferrari and A.Bilal, {\it Curves of Marginal Stability and Weak and Strong
Coupling BPS Spectra in N=2 Supersymmetric QCD,} Nucl.Phys. {\bf B480} (1996)
589, {\tt hep-th/9605101}
\semi
A.Brandhuber and S.Steiberger, {\it Self Dual Strings and Stability of BPS
States in N=2 SU(2) Gauge Theories,} Nucl.Phys. {\bf B488} (1997) 199, {\tt
hep-th/9610053}.}
\lref\rWst{N.D.Lambert and P.C.West, {\it Monopole Dynamics from the 
M-theory fivebrane,} {\tt hep-th/9811025}.}
\lref\rSpal{A. Fayyazuddin and M. Spalinski, {\it Large N superconformal gauge theories and supergravity orientifolds,} 
Nucl.Phys. {\bf B535} (1998) 219, {\tt hep-th/9805096}.}
\lref\rPS{M.K.Prasad and C.M.Sommerfeld, {\it An Exact Classical Solution
for the `t Hooft monopole and the Julia-Zee dyon,} Phys.Rev.Lett. {\bf 35}
(1975) 760.}
\lref\rGSVY{B.R. Greene, A.Shapere, C.Vafa and S.T.Yau, {\it Stringy Cosmic
Strings and Noncompact Calabi-Yau Manifolds,} Nucl. Phys. {\bf B337}
(1990) 1
\semi
M. Asano, {\it Stringy Cosmic Strings and Compactifications of F Theory,}
Nucl. Phys. {\bf B503} (1997) 177,  {\tt hep-th/9703070}.}
\lref\rnh{See for example, J.D. Bekenstein, {\it Black Hole Hair : twenty-five
years after,} {\tt gr-qc/9605059}.}
\lref\rDaction{A.A. Tseytlin, {\it Self-duality of Born-Infeld action and Dirichlet
3-brane of type IIB superstring theory,} Nucl. Phys. {\bf B469} (1996) 51, {\tt hep-th/9602064}
\semi
M. Aganic, J. Park, C. Popescu and J. Schwarz, {\it Dual D-Brane Actions,}
Nucl. Phys. {\bf B496} (1997) 215, {\tt hep-th/9702133.}}
\lref\rHarv{For a nice review see J.A. Harvey, {\it Magnetic Monopoles, Duality
and Supersymmetry,} {\tt hep-th/9603086}.}
\lref\rSols{P.Forgacs, Z.Horvath and L.Palla, {\it Nonlinear Superposition
of Monopoles,} Nucl.Phys. {\bf B192} (1981) 141
\semi
M.K.Prasad, {\it Exact Yang-Mills Higgs Monopole Solutions of arbitrary
topological charge,} Commun.Math.Phys. {\bf 80} (1981) 137.}
\lref\rSut{H.W.Braden and P.M.Sutcliffe, {\it A Monopole Metric,} Phys.Lett.
{\bf B391} (1997) 366, {\tt hep-th/9610141}
\semi
P. Sutcliffe, {\it BPS Monopoles,} Int.J.Mod.Phys. {\bf A12}
(1997) 4663, {\tt hep-th/9707009}.}
\lref\rBiel{R. Bielawski, {\it Monopoles and the Gibbons-Manton Metric,}
Commun.Math.Phys. {\bf 194} (1998) 297, {\tt hep-th/9801091}.}
\lref\rManton{N.S.Manton, {\it A Remark on the Scattering of BPS Monopoles,}
Phys.Lett. {\bf B110} (1982) 54
\semi
N.S.Manton, {\it Monopole Interactions at Long Range,} 
Phys.Lett. {\bf 154B} (1985) 395, Erratum-ibid. {\bf 157B} (1985) 475.}
\lref\rGM{G.W.Gibbons and N.S.Manton, {\it Classical and Quantum Dynamics
of BPS Monopoles,} Nucl.Phys. {\bf B274} (1986) 183
\semi
G.W.Gibbons and N.S.Manton, {\it The Moduli Space Metric for well
Separated BPS Monopoles,} Phys.Lett. {\bf B356} (1995) 32, 
{\tt hep-th/9506052}.}
\lref\rAH{M.F.Atiyah and N.J.Hitchin, {\it The Geometry and Dynamics of 
Magnetic Monopoles,} Princeton University Press, 1988.}
\lref\rSusyO{G.Gibbons, G.Papadopoulos and K.Stelle, {\it HKT and OKT 
Geometries on Soliton Black Hole Moduli Spaces,} Nucl.Phys. {\bf B508}
(1997) 623, {\tt hep-th/9706207}.}
\lref\rMSD{E. Weinberg, {\it Parameter Counting for Multi-Monopole
Solutions,} Phys.Rev. {\bf D20} (1979) 936
\semi
E. Corrigan and P. Goddard, {\it An N Monopole Solution with 4N-1
Degrees of Freedom,} Commun.Math.Phys. {\bf 80} (1981) 575.}
\lref\rGaunt{J.P. Gauntlett, {\it Low Energy Dynamics of N=2 Supersymmetric
Monopoles,} Nucl. Phys. {\bf B411} (1994) 443, {\tt hep-th/9305068}.}
\lref\rTown{J.P.Gauntlett, C.Koehl, D.Mateos, P.K.Townsend and M.Zamaklar,
{\it Finite Energy Dirac-Born-Infeld Monopoles and String Junctions,}
{\tt hep-th/9903156}.}
\lref\rDia{M.B.Green and M.Gutperle, {\it Comments on Threebranes,}
Phys.Lett. {\bf B377} (1996) 28, {\tt hep-th/9602077}
\semi
D.E.Diaconescu, {\it D-branes, Monopoles and Nahm's Equations,}
Nucl.Phys. {\bf B503} (1997) 220, {\tt hep-th/9608163}.}


\Title{ \vbox {\baselineskip 12pt\hbox{CNLS-99-03}
\hbox{BROWN-HET-1177}  \hbox{March 1999}  }}
{\vbox {\centerline{Monopole Dynamics in ${\cal N}=2$ super Yang-Mills Theory}
        \centerline{From a Threebrane Probe}
}}

\smallskip
\centerline{Robert de Mello Koch$^{1}$, 
Alastair Paulin-Campbell$^{2}$ and Jo\~ao P. Rodrigues$^{2}$}
\smallskip
\centerline{\it Department of Physics,$^{1}$}
\centerline{\it Brown University}
\centerline{\it Providence RI, 02912, USA}
\centerline{\tt robert@het.brown.edu}\bigskip

\medskip

\centerline{\it Department of Physics and Center for Nonlinear Studies,$^{2}$}
\centerline{\it University of the Witwatersrand,}
\centerline{\it Wits, 2050, South Africa}
\centerline{\tt paulin,joao@physnet.phys.wits.ac.za}\bigskip

\noindent
The BPS states of ${\cal N}=2$ super Yang-Mills theory with gauge group 
$SU(2)$ are constructed as non-trivial finite-energy solutions of the
worldvolume theory of a threebrane probe in F theory. The solutions
preserve $1/2$ of ${\cal N}=2$ supersymmetry and provide a worldvolume
realization of strings stretching from the probe to a sevenbrane. The 
positions of the sevenbranes correspond to singularities in the field 
theory moduli space and to curvature singularities in the supergravity
background. We explicitly show how the UV cut off of the effective field
theory is mapped into an IR cut off in the supergravity. Finally, we 
discuss some features of the moduli spaces of these solutions.


\Date{}


\newsec{Introduction}

Super Yang-Mills theory with gauge group $SU(N)$ and ${\cal N}=2$ supersymmetry
is realized at low energy and weak string coupling, on the worldvolume of
$N$ Dirichlet fourbranes stretched between two 
Neveu-Schwarz fivebranes\rWitten. The
singularities where the fourbranes meet the Neveu-Schwarz fivebranes are
resolved in the strong string coupling limit where this brane set up is
replaced by a single M-theory fivebrane wrapping the Seiberg-Witten curve.
In this M-theory description, the fourbranes become tubes, wrapping the
eleventh dimension and stretching between the two flat asymptotic sheets of
the M-fivebrane. The point at which they meet the fivebrane is resolved -
the fourbrane (tube) ending on a fivebrane creates a dimple in the fivebrane. 
It is a fascinating result that the non-trivial bending of the branes due
to these dimples encode the perturbative plus all instanton corrections to the
low energy effective action describing the original $SU(N)$ ${\cal N}=2$ super
Yang-Mills theory\foot{The Seiberg-Witten effective action can also be 
recovered from a Dirichlet fivebrane of IIB string theory wrapping the 
Seiberg-Witten curve\rGukov.}\rOz. The encoding of 
quantum effects in the field theory in
a bending of the brane geometry is a general feature of gauge theories
realized on brane worldvolumes. 

The M-theory description of the fivebrane is
only valid at large string coupling. Field theory however, is only recovered
at weak string coupling\rOoguri. Quantities that are protected by
supersymmetry are not sensitive to wether they are computed at weak or strong
coupling. Consequently, computing these quantities we find the fivebrane and
field theory results agree. The fivebrane does not reproduce the field theory
result for quantities that are not protected by supersymmetry. In particular,
higher derivative corrections to the low energy effective action computed
using the fivebrane do not agree with the field theory results\rOz.

An alternative approach to the study of ${\cal N}=2$ super Yang-Mills theory
is to realize these theories as the worldvolume theory of a threebrane probe in 
an F theory background\rLowe,\rOfer. The F theory 
background is set up by a collection of $N$ 
coincident threebranes, and a number of parallel F theory sevenbranes.
The sevenbranes carry R and NS charge. Our notation for the sevenbranes is
defined by saying that a $(p,q)$ string can end on a $(p,q)$
sevenbrane. Thus, for example, a (1,0) string is an elementary type IIB string;
a $(1,0)$ sevenbrane is a Dirichlet sevenbrane. In this notation, the F theory
background contains one $(0,1)$ sevenbrane, one $(2,1)$ sevenbrane and $N_f$
$(1,0)$ sevenbranes. By taking $N$ large and the string coupling small, the 
curvature of the background geometry becomes small almost everywhere and
the supergravity solution can be used reliably\rOfer. 
The field theory of interest is realized
on the worldvolume of a threebrane probe which explores 
this geometry\rLowe. On
the supergravity side we are studying the two body interaction between the
probe and the $N$ coincident threebranes. Thus the worldvolume theory of the
probe should be compared to the low energy effective action for the theory
with gauge group $SU(2)$. Higher derivative corrections to the low
energy effective action have been computed in \rRadu,\rUs 
using the probe worldvolume
theory. This work provides strong evidence that the full ${\cal N}=2$ low energy
WIlsonian action (and not just its leading Seiberg-Witten form) is described by a
Born-Infeld world-volume action of a D3 brane in the near horizon geometry of the 
above F theory background. It was shown in \rUs that the Born-Infeld action 
reproduces exactly the expected structure of non-holomorphic higher derivative
corrections to the Seiberg-Witten effective action. Essential to this approach 
is the construction of a valid background geometry, including a non trivial metric. 
Even though the leading Seiberg-Witten term is independent of the metric it turns
out to be essential for the correct description of non-holomorphic corrections.  
As is well known, ${\cal N}=2$ super Yang-Mills theories have a rich spectrum of
BPS states\rSW,\rFerrari.
It is natural to ask what the analog of the field theory BPS states on the 
threebrane worldvolume are. Sen \rSen\ 
has identified these states with strings that
stretch along geodesics from the probe to a sevenbrane. This description of
BPS states provides an elegant description of the selection rules\rSel\
determining the allowed BPS states for the $N_f\le 4$ $SU(2)$ 
Seiberg-Witten theory\rFayy. However, open string junctions can fail to be smooth\rGaber.
In this case, one needs to consider geodesic string junctions to describe the BPS states\rBerg.
\foot{We thank Oren Bergmann for helpful correspondence on this point.}
An assumption implicit in this approach is that the threebrane probe is infinitely 
heavy as compared to the string. In this approximation, the three brane remains 
flat and its geometry is unaffected by the string ending on it. Recalling the
role that bending plays in the M-theory fivebrane description of field theory,
it is natural to identify this approximation with a classical description of 
the BPS states. 

In this article, we will construct a description of BPS states of the Born-Infeld
probe world-volume action, which accounts for the bending of the threebrane into
a dimple due to the string attached to it\foot{An analogous computation for
the M-theory fivebrane has been carried out in\rWst. However, in this case the
BPS states correspond to self-dual strings stretched along the Riemann surface
associated with the Seiberg-Witten curve, instead of along the moduli space.}.
The analogy to the M-fivebrane 
description of field theory suggests that we are accounting for quantum
effects in the background of these BPS states. We are able to provide a supergravity
solution corresponding to a string stretching over a finite interval. On the 
world-volume of the D3 probe this translates into a cut-off allowing for the existence
of {\it finite energy} classical solutions of the abelian Born-Infeld action. We show that
the long distance properties of the monopole solution to non-abelian gauge theories 
are reproduced by the abelian low energy effective action. This caculation again
requires precise knowledge of the near horizon background geometry. This background 
geometry pinches the dimple on the D3 brane into an endpoint. By appealing to the holographic 
principle we are able to interpret the long distance "cut-off" on the dimple in terms of the
correct UV cut-off entering in the definition of the Wilsonian effective action. 

In section 2 we reconsider the problem of solving for the background set up
by the sevenbranes plus $N$ threebranes. This reduces to the problem of
solving the Laplace equation on the background generated by the 
sevenbranes\rOfer. We show that the solution to this Laplace equation is
duality invariant. By making an explicitly duality invariant ansatz, we are
able to accurately construct the background geometry in the large
$|a|$ region and close to the sevenbranes. A computation of the square of the
Ricci tensor shows that this background has curvature singularities. The
interpretation of these singularities in the field theory, are as the
points in moduli space where the effective action breaks down due to the appearance of new 
massless particles. 

In section 3 we construct solutions
corresponding to Dirichlet strings stretching from the probe to the
(0,1) sevenbrane. We do this explicitly for the $N_f=0$ case, but the
extension to $N_f\le 4$ is trivial.\foot{F-theory backgrounds which correspond to 
the superconformal limit of the field theories we consider have been constructed in \rSpal.} 
These "magnetic dimple" solutions have
the interesting property that they capture the long distance behaviour of
the Prasad-Sommerfeld monopole solution\rPS. 
The Prasad-Sommerfeld solution is an
exact solution for an $SU(2)$ gauge theory. An important feature of this
monopole solution, intimately connected to the non-Abelian structure of the
theory, is the way in which both the gauge field and Higgs field need to be
excited in order to obtain finite energy solutions. It is interesting that
the Abelian worldvolume theory of the probe correctly captures this structure.
This is not unexpected since the probe worldvolume theory is describing the 
low energy limit of a non-Abelian theory. The energy of these solutions is 
finite due to a cut off which must be imposed. The UV cut off in the field
theory maps into an IR cut off in the supergravity, which is a manifestation
of the IR/UV correspondence\rES.

In section four we consider the moduli space of the solutions we have 
constructed. We are able to compute the metric on the one monopole moduli
space exactly. For two monopoles and higher, 
it does not seem to be possible to
do things exactly. However, under the assumption that the monopoles are very
widely separated, we are able to construct the asymptotic form of the metric.
In this case, we see that the metric receives both perturbative and instanton
corrections. The hyper-K\"ahler structure of these moduli spaces
in unaffected by these corrections.

\newsec{Gravitational Interpretation of Singularities in the Field Theory 
Moduli Space}

The model that we consider comprises of a large number $N$ of 
coincident threebranes and a group of separated
but parallel sevenbranes. The sevenbranes have
worldvolume coordinates $(x^0,x^1,x^2,x^3,x^4,x^5,x^6,x^7)$; the 
threebranes have worldvolume coordinates $(x^0,x^1,x^2,x^3).$ The presence 
of the sevenbranes gives rise to nontrivial monodromies for the complex 
coupling\foot{The $N$ dependence of $\tau$ has been discussed in\rUs. We
will not indicate this dependence explicitly.}
$\tau=\tau_1 +i\tau_2$ as it is moved in the $(x^8,x^9)$ space transverse to
both the sevenbranes and the threebranes. In order to study ${\cal N}=2$
super Yang-Mills theory, one chooses the sevenbrane background in such a way
that the complex IIB coupling $\tau$ is equal (up to a factor) to the effective
coupling of the low energy limit of the field 
theory of interest\rUs. The spacetime
coordinates $(x^8,x^9)$ then play the dual role of coordinates in the
supergravity description, and of the (complex) Higgs field $a$ in the field
theory\foot{The precise identification between $(x^8,x^9)$ and $a$ involves a field redefinition as explained in \rJe.}. The metric due to the sevenbranes by
themselves is given by\rGSVY\

\eqn\Met
{ds^2=-(dx^0)^2+(dx^i)^2+\tau_2\big[(dx^8)^2+(dx^9)^2\big],}

\noindent
where $i$ runs over coordinates parallel to the sevenbrane.
The addition of $N$ threebranes to this pure sevenbrane 
background deforms the metric and leads to a non-zero self-dual RR fiveform
flux. This flux and the deformed metric are given in terms of $f$

\eqn\Sln
{\eqalign{ds^2 =f^{-1/2}dx_{\parallel}^{2}+f^{1/2}\big[ (dx^{i})^{2}
+&\tau_{2}\big( (dx^{8})^{2}+(dx^{9})^{2}\big)\big]=
f^{-1/2}dx_{\parallel}^2 + f^{1/2}{g}_{jk}dx^{j}dx^{k},\cr
F_{0123j}&= -{1\over 4}\partial_{j}f^{-1},}}

\noindent
where $f$ is a solution to the Laplace equation in the sevenbrane 
background\rOfer\

\eqn\LaplacEqn
{{1\over\sqrt{g}} 
\partial_{i}(\sqrt{ g}{ g}^{ij}\partial_j f)=
- (2 \pi)^4 N { \delta^6(x-x^0) \over \sqrt{\ g}. }}

\noindent
The position of the $N$ source threebranes is $x^0$.
The index $i$ in \Sln\ runs over coordinates transverse to the threebrane
but parallel to the sevenbrane, $j,k$ runs over coordinates transverse to
the threebrane and $x_{\parallel}$ denotes the coordinates
parallel to the threebrane. For the backgrounds under consideration, 
\LaplacEqn\ can be written as

\eqn\LapEqn
{\big[\tau_{2}\partial_{y}^{2}+4l_s^{-4}\partial_{a}\partial_{\bar{a}}\big]
f=-(2\pi )^{4}N\delta^{(4)}(y-y^0)\delta^{(2)}(a-a^0)}

\noindent
where $a,\bar{a}$ and $\tau_2$ are the quantities appearing in the 
Seiberg-Witten solution\rSW\ and $y$ denotes the coordinates transverse
to the threebranes but parallel to the sevenbranes.
In \rUs, this equation was solved in the large $|a|$ limit, for $\tau$ 
corresponding to ${\cal N}=2$ field theories with gauge group $SU(2)$ and
$N_f=0,4$ flavors of matter. It is possible to go beyond this approximation,
by noting that the form of \LapEqn\ implies that $f$ is invariant under
a duality transformation. To see this, change from the electric variables
$a$ to the dual magnetic variables $a_D$. After a little rewriting, we find
that \LapEqn\ becomes

\eqn\Trnfrmd
{\big[{1\over 2i}(\partial_{\bar{a}} \bar{a}_D-
\partial_a a_D )\partial_{y}^{2}
+4l_s^{-4}\partial_{a_D}\partial_{\bar{a}_D}\big]
f=-(2\pi )^{4}N\delta^{(4)}(y-y^0)\delta^{(2)}(a_D-a_D^0)}

\noindent
If we now identify $\tau_D=-{1\over \tau}$ we see that $f$ is also a solution 
of the Laplace equation written in terms of the dual variables. This result is
a consequence of the fact that the Einstein metric is invariant under the
classical $SL(2,R)$ symmetry of IIB supergravity. It is possible to construct 
an approximate solution $f$ which is valid in the region corresponding to
large $|a|$ and in the region corresponding to small $|a_D|$ 

\eqn\fsoln
{f(y,a,\bar{a})={l_s^4\over\big[y^{i}y^{i}-
{i\over 2}l_s^4(a_{D}(\bar{a}-\bar{a}_0)-\bar{a}_{D}(a-a_0))\big]^{2}}.}

\noindent
In this last formula, $a_0$ is the constant value of $a$ at $a_D=0$.
For concreteness, we will focus on the pure gauge theory in the discussion
which follows, but our results are valid for all $N_f\le 4$.
In the large $a$ limit, from the work of Seiberg and Witten\rSW, we know
that $a_D$ can be expressed as a function of $a$ as (see for example
\rLerche)

\eqn\AdA
{a_D=\tau_0 a+{2ia\over\pi}\log\Big[{a^2\over\Lambda^2}\Big]
+{2ia\over\pi}+{a\over 2\pi i}\sum_{l=1}^\infty c_l (2-4l)\Big(
{\Lambda\over a}\Big)^{4l},}

\noindent
where $\tau_0$ is the classical coupling and $\Lambda$ is the dynamically
generated scale at which the coupling becomes strong.
Inserting this into \fsoln\ and evaluating 
the expression at $y=0$ we find the
following asymptotic behaviour for $f$

\eqn\AsymptBeh
{f\sim {1\over (a\bar{a}\log|a|)^2}.}

\noindent
This reproduces the large $|a|$ solution of \rUs. In the small $|a_D|$ limit,
$a$ can be expressed as\rLerche\

\eqn\Asymptotic
{a=\tau_{0D}a_D-{2ia_D\over 4\pi}\log\Big[{a_D\over\Lambda}\Big]
-{ia_D\over 4\pi}-{1\over 2\pi ia_D}\Lambda^2\sum_{l=1}^\infty
c_l^D l\Big({ia_D\over \Lambda}\Big)^{l}.}

\noindent
Thus, in this limit and at $y=0$ we find

\eqn\LimitForF
{f\sim {1\over (\bar{a}_D a_D\log|a_D|)^2}.}

\noindent
It is easy to again check that this is the correct leading behaviour for $f$
close to the monopole singularity at $|a_D|=0$. The corrections to $f$, at
$y=0$ are of
order $|a_D|^{-4}(\log |a_D|)^{-4}$. In the large $|a|$ limit, $f\to 0$ and 
in the small $|a_D|$ limit, $f\to\infty$ signaling 
potential singularities in both limits. From
the field theory point of view, both of these limits correspond to weakly
coupled limits of the field theory (or its dual) which leads us to suspect
that curvature corrections may not be small in these regions\rWit. 
This is easily confirmed by computing the square of the Ricci tensor in 
string fame, which for large $|a|$ behaves as $R^{MN}R_{MN}\sim \log|a|$
and for small $|a_D|$ behaves as 
$R^{MN}R_{MN}\sim \log|a_D|.$
It is clear that the point $|a_D|=0$ corresponds to a
curvature singularity in the supergravity background. If we circle this
point in the field theory moduli space, the coupling transforms with a
nontrivial monodromy, so that this point is to be identified with the
position of a sevenbranes. The appearance of these naked singularities in
the supergravity background could have been anticipated from no-hair
theorems\rnh.

Recently a number of interesting type IIB supergravity backgrounds were 
constructed\rGubser. These backgrounds exhibit confinement and a running
coupling. In addition, a naked singularity appears in spacetime. It is
extremely interesting to determine whether this naked singularity can be
attributed to a weakly coupled dual description of the theory, since this
would provide an example of duality in a non-supersymmetric confining gauge
theory.

\newsec{Magnetic Dimples in the Probe Worldvolume}

The BPS states of the field theory living on a probe which explores the
sevenbrane background, have been identified with strings that stretch along
geodesics, from the probe to a sevenbrane\rSen.\foot{As mentioned in the introduction
a large number of BPS states of the field theory do not correspond to strings stretched
between the threebrane and a sevenbrane, but rather to string junctions (or webs) which 
connect the threebrane to more than one sevenbrane \rGaber,\rBerg.}  In this section
we will look for a description of these 
BPS states of the probe worldvolume, which
accounts for the bending of the threebrane into a dimple, due to the string
attached to it. We will focus on the case when a Dirichlet string ends on the 
probe, corresponding to a "magnetic dimple".  

The dynamics of the threebrane probe is given by the Born-Infeld action for a
threebrane in the background geometry of the sevenbranes and $N$ threebranes.
This background has non-zero RR five-form flux, axion, dilaton and metric.
The worldvolume action is computed using the induced (worldspace) metric 

\eqn\IndMet
{g_{mn}=G_{MN}\partial_m X^M\partial_n X^N+l_s^2 F_{mn}}

\noindent 
where $F_{mn}$ is the worldvolume field strength tensor.
We will use capital letters to denote spacetime coordinates ($X^M$) and lower
case letters for worldvolume coordinates ($x^n$). We are using the static
gauge $(X^0,X^1,X^2,X^3)=(x^0,x^1,x^2,x^3)$. In addition to the Born-Infeld
term the action includes a Wess-Zumino-Witten coupling to the 
background RR five-form field strength. The probe is taken parallel to the
stack of $N$ threebranes so that no further supersymmetry is broken when the
probe is introduced. The worldvolume soliton solutions that we will construct
have vanishing instanton number density so that there is no coupling to the
axion. The explicit action that we will use is\rDaction\

\eqn\explctact
{S=T_3\int d^4 x\sqrt{-\det (g_{mn}+e^{-{1\over 2}\phi}l_s^2F_{mn})}
+T_3\int d^4 x\partial_{n_1}X^{N_1}
\wedge ...\wedge\partial_{n_4}X^{N_4} A_{N_1 N_2 N_3 N_4},} 

\noindent
where $A_{N_1 N_2 N_3 N_4}$ is the potential for the self-dual fiveform and
we have omitted the fermions. We work in the Einstein frame so that the
threebrane tension $T_3=l_s^{-4}$ is independent of the string coupling.
In what follows, we will use $a_D$ to denote the magnetic variable that 
provides the correct description of the field theory in the small $|a_D|$ 
regime. The corresponding low energy effective coupling is denoted $\tau_D$.
The electric variable $a$ is the correct variable to use in the large $|a|$
limit and the associated coupling is denoted $\tau$.

\subsec{Spherically Symmetric Solution}

In this section we will construct a worldvolume soliton that 
can be interpreted as a single Dirichlet string 
stretching from the probe to the "magnetic" (0,1) seven brane. We
will study the pure gauge theory. The extension to $0<N_f\le 4$ is 
straightforward and amounts the same computation with a different $\tau$. From
the point of view of the worldvolume, the Dirichlet string endpoint behaves as
a magnetic monopole\rDia. If we consider 
the situation in which the probe and the
magnetic sevenbrane have the same $x^9$ coordinate and are separated in the
$x^8$ direction, then the only Higgs field which is excited is $x^8$. In 
addition, because we expect our worldvolume soliton is a magnetic monopole
we make the spherically symmetric ansatz $x^8=x^8(r)$ and assume that the
only non-zero component of the worldvolume field strength tensor in 
$F_{\theta\phi}$. With this ansatz, the threebrane action takes the form

\eqn\tbact
{S=T_3\int dtdrd\theta d\phi
\Big(\sqrt{f^{-1}+\tau_{2D}(\partial_{r}x^{8})^2}
\sqrt{f^{-1}r^4\sin^2\theta
-\tau_{2D}l_s^4 F_{\theta\phi}F_{\phi\theta}}
-{1\over f}r^{2}\sin\theta\Big).}

\noindent
We will consider the situation in which the probe is close to the (0,1)
sevenbrane corresponding to a region where $a_D$ is small. For this reason
we use the dual coupling in \tbact\ and we will identify $x^8$ with the
dual Higgs $a_D$.  
Recall the fact that the Born-Infeld action has the property that a BPS
configuration of its Maxwellian truncation satisfies the equation of motion
of the full Born-Infeld action\rMal. 
In our case, the Maxwellian truncation of
the Born-Infeld action is just Seiberg-Witten theory. 
The supersymmetric
variation of the gaugino of the Maxwell theory is

\eqn\MT
{\delta \psi=(\Gamma_{\theta\phi}F^{\theta\phi}
+\Gamma_{8r}\partial_{r}x^8)\epsilon.}

\noindent
A BPS background is one for which this variation vanishes. Now, note that if
we identify ($\eta^{mn}$ is the flat four dimensional Minkowski metric)

\eqn\BPSCond
{\eta^{\theta\theta}\eta^{\phi\phi}l_s^4(F_{\theta\phi})^2=
{F_{\theta\phi}^2l_s^4\over r^4\sin^2\theta}=(\partial_r x^8)^2=
\eta^{rr}(\partial_r x^8)^2,}

\noindent
we obtain a background invariant under supersymmetries \MT, where $\epsilon$
satisfies $(\hat{e}^{\theta}\Gamma_{\theta}\hat{e}^{\phi}\Gamma_{\phi}$
$\pm\hat{e}^8\Gamma_{8}\hat{e}^r\Gamma_{r})\epsilon =0.$ The sign depends on
wether we consider a monopole or an anti-monopole background. This
condition can be rewritten as
$\Gamma_{1}\Gamma_{2}\Gamma_{3}\Gamma_{8}\epsilon=\pm\epsilon$.
Thus, this is a BPS background of Seiberg-Witten theory. We will assume 
that \BPSCond\ provides a valid BPS condition for the full Born-Infeld action.
Upon making this ansatz, we find that the determinant factor in the Born-Infeld
action can be written as a perfect square, as 
expected\rHash. The full Born-Infeld
action plus Wess-Zumino-Witten term then simplifies to the Seiberg-Witten
low energy effective action \foot{The study of the quantum corrections to solitons by 
studying the minima of the low energy effective action has been considered in \rChalm.}

\eqn\SWAct
{S=T_3\int d^4 x\tau_{2D}(\partial_{r}x^{8})^2.}

\noindent
We have checked that the arguments above can also be made at the level of the 
equations of motion. The preserved supersymmetries have a natural 
interpretation. The probe preserves supersymmetries of the form
$\epsilon_L Q_L+\epsilon_R Q_R$ where
$\Gamma_0\Gamma_1\Gamma_2\Gamma_3\epsilon_L=\epsilon_R$. The Dirichlet
string preserves supersymmetries for which
$\Gamma_0\Gamma_8\epsilon_L=\pm\epsilon_R$, with the sign depending on wether
the string is parallel or anti-parallel to the $x^8$ axis. The two
conditions taken together imply that 
$\Gamma_1\Gamma_2\Gamma_3\Gamma_8\epsilon_i=\pm\epsilon_i$, $i=L,R$. Thus,
the supersymmetries preserved by the ansatz \BPSCond\ are exactly the 
supersymmetries that one would expect to be preserved by a Dirichlet string
stretched along the $x^8$ axis. With a suitable choice of the phase of $a_D$
we can choose \foot{This identification is only correct to leading 
order at low energy. One needs to perform a field redefinition of the field theory Higgs fields before they can be identified with supergravity 
coordinates\rJe.} $a_Dl_s^2=-ix^8$. With this choice, 
using the formulas for the dual prepotential
quoted in \rLerche, we find that the dual coupling 
can be expressed in terms of
$x^8$ as 

\eqn\DulCoup
{\tau_{2D}={4\pi\over g_{cl,D}^2}
-{3\over 4\pi}-{1\over 2\pi}\log\Big[{x^8\over l_s^2\Lambda}\Big]
-{1\over 2\pi}\Lambda^2l_s^4\sum_{l=1}^\infty {c_l^D l(l-1) (x^8)^{l-2}\over
(l_s^2\Lambda)^l}.}

\noindent
Note also that $a$ is real and can be expressed in terms of $x^8$ and 
$\tau_{2D}$
as

\eqn\ExpFrA
{a=\int dx^8 {\tau_{2D}\over l_s^2}={4\pi\over g_{cl,D}^2}{x^8\over l_s^2}
-{1\over 4\pi}{x^8\over l_s^2}
-{1\over 2\pi l_s^2}x^8\log\Big[{x^8\over l_s^2\Lambda}\Big]
-{1\over 2\pi}\Lambda^2l_s^2\sum_{l=1}^\infty {c_l^D l (x^8)^{l-1}\over
(l_s^2\Lambda)^l},}

\noindent
The equation of motion which follows from \SWAct\

\eqn\SWEom
{{d\over d r}\Big(\tau_{2D} r^2{dx^8\over dr}\Big)=0,}

\noindent
is easily solved to give

\eqn\Dimple
{\gamma-{\alpha\over r}=a={4\pi\over g_{cl,D}^2 l_s^2}x^8
-{1\over 4\pi l_s^2}x^8
-{1\over 2\pi l_s^2}x^8\log\Big[{x^8\over l_s^2\Lambda}\Big]
-{1\over 2\pi}\Lambda^2l_s^2 \sum_{l=1}^\infty 
{c_l^D l (x^8)^{l-1}\over (l_s^2\Lambda)^l}.}

\noindent
In principle, this last equation determines the exact profile of our magnetic
dimple as a function of $r$. At large $r$ our fields have the following
behaviour

\eqn\Asympt
{{x^8\over l_s^2}=\nu -{\beta\over r},\quad
l_s^4 B^{r}B_r=l_s^4 F^{\theta\phi}F_{\theta\phi}=(\partial_r x^8)^2=
{\beta^2 l_s^4\over r^4},}

\noindent
where $\nu$ and $\beta$ are constants.
To interpret these results, it is useful to recall the Prasad-Sommerfeld
magnetic monopole solution\rPS. This monopole is a solution to the following
$SU(2)$ non-Abelian gauge theory

\eqn\Nalbi
{\eqalign{S&=-{1\over e^2}\int d^4 x\Big({1\over 4}(F^a_{mn})^2+
{1\over 2}(D^n\phi^a)^2\Big),\cr
F^a_{mn}=\partial_m A_n^a&-\partial_n A_m^a +\epsilon^{abc}
A_m^bA_n^c,\quad D_n\phi^a=\partial_n\phi^a+\epsilon^{abc}
A_n^b\phi^c.}}

\noindent
An important feature of the monopole solution, intimately connected to
the non-Abelian gauge structure of the theory, is the fact that in order
to get a finite energy solution, the monopole
solution excites both the gauge field and the Higgs field\rHarv. The Higgs
field component of the Prasad-Sommerfeld monopole is

\eqn\PSMonopole
{\eqalign{\phi^a
&=\hat{r}^a\Big(\gamma \coth ({\gamma r\over\alpha})
-{\alpha\over r}\Big),\cr
&=\gamma -{\alpha\over r}+2\gamma e^{-2\gamma r\over\alpha}+2\gamma
e^{-4\gamma r\over\alpha}...,}}

\noindent
where on the second line we have performed a large $r$ expansion. Clearly,
the electric variable\foot{It is the electric variable $a$ of
Seiberg-Witten theory that is related to the Higgs field appearing in the
original microscopic $SU(2)$ theory.} $a$ reproduces the large $r$ behaviour
of the Higgs field of the Prasad-Sommerfeld solution. Thus, we see that
the Abelian worldvolume theory of the threebrane probe catches some
of the structure of the non-Abelian field theory whose low energy limit
it describes. Comparing $a$ to the asymptotic form of the Higgs field in the
Prasad-Sommerfeld solution allows us to interpret the constants of integration
$\alpha$ and $\gamma$. The constant $\alpha$ is related to the inverse of
the electric charge $1/e$. As $r\to\infty$ $a\to\gamma$ so that $\gamma$ 
determines the asymptotic Higgs expectation value; i.e. it determines the 
moduli parameters $a,\bar{a}$ of the field theory or, equivalently, the 
position of the threebrane in the $(8,9)$ plane. The mass of the $W^\pm$
bosons are given by the ratio $m_W=\gamma/\alpha.$ In a similar way, $\nu$
fixes the asymptotic expectation value of the dual Higgs field and $\beta$
is related to the inverse magnetic charge $1/g$. Thus, the mass of the BPS
monopole is given by $m_g=\nu/\beta.$ We are working in a region of moduli
space where $|a_D|$ is small, so that the monopoles are lighter that the
$W^\pm$ bosons. We will return to the exponential corrections in \PSMonopole\
below.

Consider next the small $r$ limit. It is clear that in this limit 
$a\to\infty$. To correctly interpret this divergence in $a$, recall that
the Seiberg-Witten effective action is a Wilsonian effective action, obtained
by integrating out all fluctuations above the scale set by the mass of
the lightest BPS state in the theory. In our case, the lightest BPS states 
are the monopoles and this scale is $m_g$. 
The largest fluctuations left in the effective low energy
theory all have energies less than $m_g$. By Heisenberg's uncertainty
relation, the effective theory must be cut off at a smallest distance of
$1/m_g$. Clearly then, the divergence in the $r\to 0$ limit is unphysical and
it occurs at length scales below which the effective theory is valid. What is
the interpretation of this short distance (UV) field theory cut off in the
supergravity description? To answer this question, we will need a better 
understanding of the small $r$ behaviour of $x^8$. Towards this end, note that
with our choice $a_D$, $\tau$ is pure imaginary so that

\eqn\Chad
{\tau=-{1\over\tau_D}={i\over\tau_{2D}}=i\tau_2.}
 
\noindent
Thus, the action \SWAct\ can be written as

\eqn\NwAct
{S=T_3\int d^4 x\tau_2 (\partial_r a)^2=
T_3\int d^4 x {(\partial_r a)^2\over\tau_{2D}}.}

\noindent
Noting that $\tau_{2D}=\partial a/\partial x^8$, the equation of motion 
for $a$ implies 

\eqn\Crnch
{r^2\tau_2{\partial a\over\partial r}=r^2{\partial x^8\over \partial r}}

\noindent
is a constant. Thus, the expressions in \Asympt\ are exact.
At $r=1/m_g=\nu/\beta$ we find that

$$ a_D=-i\Big(\nu-{\beta\over r}\Big)=0.$$

\noindent
Thus, the magnetic dimple is cut off at $a_D=0$. Intuitively this is pleasing:
the magnetic dimple (Dirichlet string) should end on the 
magnetic (0,1) sevenbrane
which is indeed located at $a_D=0$. To understand the geometry of the dimple
close to $a_D=0$, note that the induced metric is

\eqn\WorldMet
{\eqalign{ds^2&=f^{-1/2}(-dt^2+d\Omega_2)
+(f^{-1/2}+f^{1/2}\tau_2\partial_r x^8\partial_r x^8)dr^2\cr
&=l_s^2 a_D\bar{a}_D\log|a_Dl_s| (-dt^2+d\Omega_2)+\cr
&(l_s^2 a_D\bar{a}_D\log|a_Dl_s|+
{\tau_{2D} \over l_s^2 a_D\bar{a}_D\log|a_Dl_s|}
\partial_r x^8\partial_r x^8)dr^2.}}

\noindent
In the above expression we have used the approximate solution \fsoln\
for the metric, obtained in the last section. This is a valid
approximation since we are interested in the geometry close to $a_D=0$
where our expression for $f$ becomes exact. The proper length to $a_D=0$ is
clearly infinite. It is also clear from \WorldMet\ that the dimple ends at
a single point at $a_D=0$, which is again 
satisfying. Thus, although we motivated
the need to introduce a cut off from the point of view of the low energy
effective action realized on the probe, one could equally well 
argue for {\it exactly
the same} cut off from the dual gravity description. 
In the super Yang-Mills theory 
one has a short distance (UV) cut off; on the gravity side one has a long
distance (IR) cut off. The connection between these cut offs is expected
as a consequence of the UV/IR correspondence.

A direct consequence of the short distance cut off in field theory, is that
the monopole appears as a sphere of radius $1/m_g$. From the point of view of 
the effective field theory, it is not possible to localize the monopole any
further. The more a particle in quantum field theory is localized, the higher
the energy of the cloud of virtual particle fluctuations surrounding it will 
be. If one localizes the monopole in the effective field theory any further,
the energy of the fluctuations becomes high enough to excite virtual 
monopole-antimonopole 
pairs, and hence we would leave the domain of validity of the effective
field theory. The origin of the exponential corrections in \PSMonopole\ can
be traced back to virtual $W^\pm$ bosons by noting that the factors multiplying
$r$ in the exponent are proportional to the boson mass 
$m_W$. The Higgs field of the 
Prasad-Sommerfeld monopole goes smoothly to zero as $r\to 0$ so that these 
bosons resolve the singular monopole core. The fact that these fluctuations have
been integrated out of the effective theory naturally explains why $a$ does
not receive any exponential corrections. Although these corrections are crucial
for the description of the monopole core, they are not needed by the effective 
field theory which describes only low energy (large distance) phenomena. In a
similar way, virtual monopole-antimonopole pairs will resolve the singular
behaviour of $x^8$ as $r\to 0$.

The cut-off that we employ in our work has already been anticipated in a completely 
different context in \foot{We would like to thank Frederik Denef for bringing this to
our attention.} \rDenef. In \rDenef the effective field theory realisation of BPS states
was studied in the field theory limit of IIB string theory compactified on a Calabi-Yau
manifold. In this description, the BPS states arise as limits of the attractor "black holes" 
in ${\cal N}=2$ supergravity. To analyse the BPS equations around $a_D=0$ (a singular point 
in the field theory moduli space), it is safer to employ the "attractor-like" formulation of 
the BPS equations. The solution with a finite $a_D=0$ core then arises as a solution of the BPS
equations of motion. Our results are in agreement with those of \rDenef. 

We will now comment on the relation between our study and the results presented in \rWst.
Above we argued that there was a need for a cut-off because the world-volume theory of 
the probe is a Wilsonian effective action, obtained by integrating all fluctuations 
above some energy scale out of the theory. However, this argument could also be made 
directly in the supergravity description by examining the induced metric given above. 

The form of this induced metric relies crucially on the fact that the probe is moving 
in the background of sevenbranes and $N$ threebranes. In particular, by taking $N\to\infty$
the authors of \rOfer argued that the background geometry can be trusted in the field theory 
limit. By accounting for the deformation of the background due to these $N$ threebranes 
our analysis goes beyond the low energy approximation and, in particular, can be trusted when
computing quantities not protected by supersymmetry. Note that the induced metric is non-holomorphic.
In the case of the M-theory fivebrane, one does not expect to reproduce quantities that are non-holomorphic \rOz.
Indeed, the background geometry for the M-theory fivebrane analysis is flat, so it is difficult
to see how an analog of the induced metric could be realised. It is an open question as to whether 
a cut-off can consistently be used in this case.  

We now compute the energy of the magnetic dimple. Since the dimple is at rest,
this energy should be proportional to the mass of the monopole. As discussed
above, $1/\alpha$ is playing the role of the electric charge $e$. The Dirac
quantization condition is $eg=4\pi$, so that we can identify the magnetic
charge $g=4\pi\alpha.$ The magnetic dimple is an excitation of the flat
threebrane located at $a=\gamma$. We will denote the corresponding value
of the dual variable $a_D$ by $a_D^0$. Thus, the mass of a magnetic monopole
of this theory is $m=g|a_D^0|=4\pi\alpha |a_D^0|$. To compute the mass of the
magnetic dimple note that (we have set the $\theta$-angle to zero) 

$$\tau_{2D}=-i{\partial a\over\partial a_D}=-i
{\partial a\over\partial r}{\partial r\over\partial a_D}
=i{\alpha\over r^2}{\partial r\over\partial a_D}.$$

\noindent
Since the kinetic energy vanishes, the energy of the soliton is proportional
to the Lagrangian

$$\eqalign{E&=-L=-l_s^{-4}\int_{0}^{\infty} dr \int_{0}^{\pi}d\theta 
\int_{0}^{2\pi}d\phi \sin\theta r^2\tau_{2D} 
(\partial_r x^8)^2 \cr 
&=+i4\pi\alpha\int^{a_D^0}_{0} da_D=4\pi\alpha|a_D^0|.}$$

\noindent
The mass of the dimple lends further support to its interpretation as a
magnetic monopole.

\subsec{Multimonopole Solutions}

In this section we will describe solutions containing an arbitrary number of
separated static dimples. We will look for static solutions, that have
$x^8$ and $F_{ij}$ $i,j=1,2,3$ excited. It is useful 
to again consider the Maxwellian
truncation of the full Born-Infeld action to motivate an ansatz. The variation
of the gaugino of the Maxwellian theory, assuming the most general static
magnetic field, is given by

\eqn\Vart
{\delta \psi=(\Gamma_{ij}F^{ij}
+\eta^{kk}\Gamma_{8k}\partial_{k}x^8)\epsilon.}

\noindent
Thus, a solution which satisfies

\eqn\ScndAnz
{{1\over 2}\eta_{il}\epsilon^{ljk}F_{jk}=
\pm \partial_{i}x^8,}

\noindent
will be invariant under supersymmetries \Vart\ as long as $\epsilon$ satisfies
$\Gamma_1\Gamma_2\Gamma_3\Gamma_8\epsilon=\pm\epsilon.$ The choice of sign
again depends on whether one is describing a monopole or an anti-monopole
background. As explained above, these are exactly the supersymmetries that one
would expect to be preserved by a Dirichlet string stretched along the $x^8$
axis. Upon making this ansatz, we again find that the determinant appearing 
in the Born-Infeld action can be written as a perfect square and that once again
the dynamics for the field theory of the threebrane probe worldvolume is
described by the Seiberg-Witten low energy effective action. After noting that
the identification $a_Dl_s^2=-ix^8$ 
implies that $\tau_{2D}=l_s^2 \partial a/\partial x^8$,
the equation of motion following from the Seiberg-Witten effective action 
can be written as

\eqn\SWEqn
{{\partial\over\partial x^i}
{\partial\over\partial x^i} a=0.}

\noindent
This is just the free Laplace equation which is easily solved

\eqn\MultiMonSol
{a=\gamma +\sum_{i=1}^{n}{\alpha\eta\over \big[
(x-x_i)^2+(y-y_i)^2+(z-z_i)^2\big]^{1/2}},}

\noindent
where $\eta=\pm 1$.
The number of dimples $n$ and their location $(x_i,y_i,z_i)$ is completely
arbitrary. This is to be expected - BPS states do not experience a static force.
In the above, we have set each of the numerators equal to $\alpha$, because as
we have seen above this factor is related to the electric charge $e$. The 
classical solution would of course allow arbitrary coefficients. The
coefficients of these terms can be positive or negative. All strings must
have the same orientation for this to be a BPS state. Thus the interpretation
of terms with a positive coefficient is that they correspond to
strings that end on the threebrane; terms
with a negative coefficient correspond to strings that
start from the threebrane. The finite energy solutions
that we consider only allow for strings that start from the probe and end
on the sevenbrane, and consequently we fix the sign of all terms to be negative.
Each of the dimples above ends in a point at the sevenbranes at $a_D=0$. The
electric Higgs field $a$ again reproduces the known asymptotic behaviour of the
Higgs field component of multi-monopole 
solutions in non-Abelian ($SU(2)$) gauge theories\rSols.

We will now consider the mass of this multi-monopole solution.
In the limit that the dimples are very widely separated, the mass
of each dimple can be computed separately. 
In this widely separated dimple limit, the
total energy is simply the sum of the energy for each dimple, so that the
energy of the $n$-dimple solution is indeed consistent with its interpretation as an $n$-monopole state. Ofcourse the total energy of the $n$-dimple state
is independent of the locations of each dimple so that once the energy is
known for widely separated dimples, it is known for all possible positions of
the dimples.
 
\newsec{Metric of the Monopole Moduli Space}

In the previous section we constructed smooth finite energy solutions to the
equations of motion. The solution describing $n$ BPS monopoles is known to be 
a function of $4n$ moduli parameters\rMSD. 
The position of each monopole accounts 
for $3n$ of these parameters. The remaining $n$ parameters correspond to
phases of the particles. By allowing these moduli to become time dependent
it is possible to construct the low energy equations of motion for the 
solitons\rManton. The resulting low energy 
monopole dynamics is given by finding
the geodesics on the monopole moduli space. The low energy dynamics of
${\cal N}=2$ supersymmetric monopoles has been studied in \rGaunt.
In this section we would like
to see if the metric on monopole moduli space can be extracted from the
solutions we constructed above.

\subsec{One Monopole Moduli Space}

A single monopole has four moduli parameters. Three of these parameters may
be identified with the center of mass position of the monopole, $x^i$, whilst
the fourth moduli parameter corresponds to the phase of the monopole, $\chi$. 
We will denote these moduli as $z^{\alpha}=(\chi,x^1,x^2,x^3)$. After allowing
the moduli to pick up a time dependence, the action picks up the following
additional kinetic term

\eqn\AddKin
{S=\int d^4 x\tau_{2D} \dot{a}_D\dot{\bar{a}}_D=\int dt 
{\cal G}_{\alpha\beta}\dot{z}^{\alpha}
\dot{z}^{\beta},}

\noindent
where we have introduced the metric on the one monopole moduli space

\eqn\OneMonMet
{{\cal G}_{\alpha\beta}\equiv \int d^3 x\tau_{2D}
{\partial a_D\over\partial z^{\alpha}} 
{\partial \bar{a}_D\over\partial z^{\beta}}.}

\noindent
Consider first the center of mass of the monopole. To construct the moduli space
dependence on these coordinates, replace $x^i_0\to x^i_0+x^i(t)$, where $x^i_0$
denotes the initial (time independent) monopole position. The action
picks up the following kinetic terms

\eqn\AdKn
{S=\int d^4 x \tau_{2D}  \dot{a}_D\dot{\bar{a}}_D=
\int d^4 x \tau_{2D} {\partial a_D\over \partial x^{i}(t)}
{\partial \bar{a}_D\over \partial x^{j}(t)}\dot{x}^{i}(t)
\dot{x}^{j}(t)=\int dt \dot{x}^{i}(t)
\dot{x}^{j}(t){\cal G}_{ij}.}

\noindent
Since the velocity of the monopole is small, we keep only terms which
are quadratic in the monopole velocity. Thus, when we compute

\eqn\OneMonMet
{{\cal G}_{ij}=-\int d^3 x\tau_{2D} {\partial a_D\over\partial x^i}
{\partial a_D\over\partial x^j}=-\int drd\theta d\phi r^2\sin \theta\tau_{2D}
\Big({\partial a_D\over\partial r}\Big)^2{\partial r\over\partial x^i}
{\partial r\over\partial x^j}}

\noindent
we evaluate the integrand at the static monopole solution. Using the
result 

$$ \int_0^{\pi} d\theta\int_{0}^{2\pi} d\phi 
\sin (\theta) {\partial r\over\partial x^i}
{\partial r\over\partial x^j}={1\over 2}4\pi\delta_{ij},$$

\noindent
we find

\eqn\ReltsFrMet
{{\cal G}_{ij}=-{1\over 2}\delta_{ij}\int dr r^2
\tau_{2D}\Big({\partial a_D\over\partial r}\Big)^2={1\over 2}
4\pi\alpha|a_D^0|
\delta_{ij}.}

\noindent
Consider now the fourth moduli parameter. Under the large gauge transformation
$g=e^{\chi(t) a_D}$ one finds that

\eqn\LGT
{\delta A_i =\partial_{i}(\chi (t) a_D),\quad 
   \delta A_0= \partial_{0}(\chi (t) a_D),\quad
   \delta a_D=0. }

\noindent
Note that since the gauge group $U(1)$ is compact, the parameter $\chi (t)$ is
a periodic coordinate. It is possible to modify this transformation so that
the potential energy remains constant and a small electric field is turned on. 
The modified transformation is\rHarv\

\eqn\Modfd
{\delta A_i =\partial_{i}(\chi (t) a_D),\quad 
   \delta A_0= \partial_{0}(\chi (t) a_D)-\dot{\chi}a_D,\quad
   \delta a_D=0.}

\noindent
After the transformation, the electric field is 
$E_i=F_{0i}=i\dot{\chi}\partial_i a_D=\dot{\chi}B_i.$ The fourth parameter is
$\chi$ and its velocity controls the magnitude of the electric field which is
switched on. The extra kinetic contribution to the action is

\eqn\LstKin
{S={1\over 2}\int d^4 x F_{0r}^2=
{1\over 2}\int dt \dot{\chi}^2\int d^3 x B_{r}B^{r}=
{1\over 2}\int dt \dot{\chi}^2 4\pi\alpha |a_D^0|. }

\noindent
Thus, we find that the quantum mechanics for the collective coordinates on 
the one monopole moduli space is described by the action

\eqn\FnlRslt
{S=\int dt {1\over 2}4\pi\alpha
|a_{D}^0|\delta_{\alpha\beta}z^{\alpha}z^{\beta}.}

\noindent
This is the correct result\rGM. Thus, 
the monopole moduli space is topologically
$R^3\times S^1$. The induced metric on the moduli space is simply the flat
metric. It is well known that this metric is hyper-K\"ahler. This fits nicely
with the structure of the moduli space quantum mechanics: the dimple
preserves ${\cal N}=1$ supersymmetry in the four dimensional field theory.
As a result, we would expect an action with ${\cal N}=4$ worldline 
supersymmetry. In one dimension there are two types of multiplets with
four supercharges. The dimensional reduction of two dimensional $(2,2)$
supersymmetry leads to ${\cal N}=4A$ supersymmetry. The presence of this
supersymmetry requires that the moduli space be a 
K\"ahler manifold\rSusyO. The 
second supersymmetry, ${\cal N}=4B$ is obtained by reducing two dimensional
$(4,0)$ supersymmetry. The presence of this supersymmetry requires that the
moduli space is a hyper-K\"ahler manifold\rSusyO.
The only fermion zero modes in
the monopole background in ${\cal N}=2$ super Yang-Mills theory is chiral  
in the sense $\Gamma_1\Gamma_2\Gamma_3\Gamma_8\epsilon=\pm\epsilon$ as
explained above. Thus, the supersymmetry on the worldline is 
${\cal N}=4B$ supersymmetry. The fact that the moduli space is 
hyper-K\"ahler is a direct consequence of supersymmetry. 

\subsec{Multi Monopole Moduli Space}

The asymptotic metric on the moduli space of 
two widely separated BPS monopoles
has been constructed by Manton\rManton. In this approach, one considers
the dynamics of two dyons. After constructing the Lagrangian that describes
the dyons motion in $R^3$, with constant electric charges as parameters, one
identifies the electric charges as arising from the motion on circles
associated with the fourth moduli parameter of the monopoles. The explicit form
of the metric for the relative collective coordinates is\rManton

\eqn\GMMet
{ds^2=U(r)dr^idr^i+{g^2\over 2\pi MU(r)}(d\chi+\omega^idr^i)^2,\quad
U(r)=1-{g^2\over 2\pi M(r^jr^j)^{1/2}},}

\noindent
where $r^i$ is the relative coordinate of the two monopoles, $\chi$ is the
relative phase, $M$ the monopole mass, $g$ the magnetic charge and $\omega^i$
the Dirac monopole potential which satisfies 
$\epsilon^{ijk}\partial^j\omega^k=r^i/(r^jr^j)^{3/2}.$
This is just the Taub-NUT metric with negative mass.
In this section, we will show that it is possible to reproduce the first term
in this metric from the dimple solutions. The second term could presumably be
reproduced by studying dyonic dimples.

We will consider the following two-dimple solution

\eqn\MDSol
{a=\gamma -\sum_{i=1}^{2}{\alpha\over |\vec{x}-\vec{x}_{i}|}.} 

\noindent
In the case of the two dimple solution, it is no longer possible to compute 
the moduli space metric exactly, and we have to resort to approximate
techniques. To reproduce the first term in \GMMet\ we need to evaluate

\eqn\IntToEval
{{\cal G}_{\alpha\beta}=-\int d^3 x\tau_{2D}
{\partial a_D\over\partial x^\alpha}
{\partial a_D\over\partial x^\beta}=
i\int d^3 x
{\partial a_D\over\partial x^\alpha}
{\partial a\over\partial x^\beta},}

\noindent
where $\alpha,\beta$ can take any one of six values corresponding to any of
the three spatial coordinates of either monopole. We do not know $a_D$ as a
function of the $x^\alpha$; we will write \MDSol\ as $a=\gamma-\Delta $. 
Setting $a_D= a_D^0+\beta_1\Delta+O(\Delta^2)$ in the expression
for $a(a_D)$ we find

\eqn\Expnd
{a(a_D)=a(a_D^0)+\beta_1\Delta{\partial a\over\partial a_D}|_{a_D=a_D^0}
=\gamma-\Delta.}

\noindent
This approximation is excellent at large $r$ where $\Delta<<1$.
Thus, we find that 

\eqn\Expnded
{a_D=a_D^0-\Delta\left({\partial a\over\partial a_D}|_{a_D=a_D^0}\right)^{-1}
=a_D^0+i{\Delta\over\tau_{2D}(a_D^0)}.}

\noindent
The dual theory is weakly coupled, so that $\tau_D(a_D^0)$ is large. Thus,
the correction to $a_D^0$ in the expression for $a_D$ is indeed small and
the approximation that we are using is valid. Using this expression for
$a_D$ in \IntToEval\ we find

\eqn\NextStep
{{\cal G}_{ij}=\int d^3 x{\alpha^2\over\tau_{2D}(a_D^0)}
\left({\partial\over\partial x^i_1}{1\over |\vec{x}-\vec{x}_1|}\right)
\left({\partial\over\partial x^j_1}{1\over |\vec{x}-\vec{x}_1|}\right).}

\noindent
To evaluate this integral it is useful to change coordinates to a spherical
coordinate system centered about monopole 1. In these coordinates

\eqn\NStep
{{\cal G}_{ij}=\delta_{ij}{1\over 2}{4\pi\alpha^2\over\tau_{2D}(a_D^0)}
\int dr{1\over r^2}.}

\noindent
The integral must again be cut off at the lower limit\foot{Strictly speaking
we should also exclude a circular region centered around
$\vec{x}=\vec{x}_1-\vec{x}_2$. However, the integrand is of order 
$|\vec{x}_1-\vec{x}_2|^{-2}$ in this region and in addition the area of the
region to be excluded is $\pi/m_g^2$, so that this is a negligible effect.}
where $a_D=0$. The value of the cut off is given by

\eqn\CtOff
{\eqalign{0=&a_D^0+{i\alpha\over \tau_{2D}(a_D^0)r}+
{i\alpha\over \tau_{2D}(a_D^0)|\vec{x}+\vec{x}_1-\vec{x}_2|}
=a_D^0+{i\alpha\over \tau_{2D}(a_D^0)r}+
{i\alpha\over \tau_{2D}(a_D^0)r_{12}}
+O({r\over r_{12}^2})\cr
{1\over r}&={\tau_{2D}(a_D^0)|a_D^0|\over\alpha}
-{1\over r_{12}},}}

\noindent
where $r_{12}$, the magnitude of the relative coordinate, is assumed to be
large. Thus, we find that

\eqn\Resltt
{{\cal G}_{ij}={1\over 2}\delta_{ij}4\pi {\alpha^2\over\tau_{2D}(a_D^0)}
\Big({\tau_{2D}(a_D^0)|a_D^0|\over\alpha}
-{1\over r_{12}}\Big)={1\over 2}\delta_{ij}
\left(4\pi\alpha|a_D^0|-
{g^2\over 4\pi\tau_{2D}(a_D^0)r_{12}}\right).}

\noindent
In a similar way, we compute

\eqn\NextSep
{\eqalign{{\cal G}_{i+3,j+3}&=\int d^3 x{\alpha^2\over\tau_{2D}(a_D^0)}
\left({\partial\over\partial x^i_2}{1\over |\vec{x}-\vec{x}_2}|\right)
\left({\partial\over\partial x^j_2}{1\over |\vec{x}-\vec{x}_2}|\right)\cr
&={1\over 2}\delta_{ij}
\left(4\pi\alpha|a_D^0|-
{g^2\over 4\pi\tau_{2D}(a_D^0)r_{12}}\right).}}

\noindent
To finish the calculation of the metric on the two monopole moduli space,
we need to compute

\eqn\NtStep
{{\cal G}_{i+3,j}=\int d^3 x{\alpha^2\over\tau_{2D}(a_D^0)}
\left({\partial\over\partial x^i_2}{1\over |\vec{x}-\vec{x}_2|}\right)
\left({\partial\over\partial x^j_1}{1\over |\vec{x}-\vec{x}_1|}\right).}

\noindent
We do not need to evaluate \NtStep\ directly. If we introduce center of mass
and relative coordinates as

\eqn\Comrel
{\vec{r}_{cm}={1\over 2}(\vec{x}_1+\vec{x}_2),\quad
\vec{r}_{12}=\vec{x}_1-\vec{x}_2,}

\noindent
it is a simple exercise to compute the center of mass contribution
to the action

\eqn\COMcont
{S=\int dt\dot{r}_{cm}^i\dot{r}_{cm}^i 4\pi\alpha |a_D^0|.}

\noindent
The integral that had to be performed to obtain this result was
proportional to the action itself. This result fixes

\eqn\LstRlt
{{\cal G}_{i+3,j}={\cal G}_{i,j+3}=
{1\over 2}\delta_{ij} 
{g^2\over 4\pi\tau_{2D}(a_D^0)r_{12}}.}

\noindent
Putting the above results together, we find the action which governs
the relative motion of the two monopoles is

\eqn\RelMot
{S_{rel}=\int dt\Big( {4\pi\alpha |a_D^0|\over 4}-
{g^2\over 8\pi\tau_{2D}(a_D^0) r_{12}}\Big)
{dr_{12}^i\over dt}{dr_{12}^i\over dt}.} 

\noindent
Thus, the low energy relative motion is geodesic motion for a metric on $R^{3}$
given by $ ds^2= U(r)dr^i dr^i,$ with 
$U(r)=1-g^2/(8\pi^2\alpha |a_D^0|\tau_{2D}(a_D^0) r)$. This
reproduces the first term in \GMMet. Notice however that the magnetic coupling
comes with a factor of $1/\tau_{2D}(a_D^0)$. This 
factor has its origin in the loop plus
instanton corrections that were summed to obtain the low energy effective
action. These corrections do not change the fact that the two monopole moduli
space is hyper-K\"ahler.
Thus, the dimple solutions on the probe reproduces the quantum corrected
metric. This non-trivial metric has its origin in the fact that the forces due
to dilaton and photon exchange no longer cancel at non-zero velocity due to the
different retardation effects for spin zero and spin one exchange\rHarv.

The exact two monopole metric has been determined by Atiyah and Hitchin\rAH. 
The fact that it has an $SO(3)$ isometry arising from rotational invariance, 
that in four dimensions hyper-K\"ahler implies self-dual curvature and the fact
that the metric is known to be complete determines it exactly. Expanding the
Atiyah-Hitchin metric and neglecting exponential corrections, one recovers the
Taub-NUT metric\rGM. If we again 
interpret the origin of the exponential corrections
as having to do with virtual $W^\pm$ boson effects, it is natural to expect
that the exact treatment of the dimples in the probe worldvolume theory will
recover the (quantum corrected) Taub-NUT metric and not the Atiyah-Hitchin 
metric. The metric on the moduli space of $n$ well separated monopoles has been
computed by Gibbons and Manton\rGM\ by studying 
the dynamics of $n$ well separated
dyons. The exact monopole metric computed for a tetrahedrally symmetric charge
4 monopole was found to be exponentially 
close to the Gibbons-Manton metric\rSut.
This result was extended in \rBiel\ where it was shown that the Gibbons-Manton
metric is exponentially close to the exact metric for the general $n$
monopole solution. In view of these results it is natural to conjecture that
an exact treatment of the $n$-dimple solution in the probe worldvolume theory
will recover the (quantum corrected) Gibbons-Manton metric. 

{\it Note Added:} When this work was near completion, we received \rTown.
In this work finite energy BPS states corresponding to open strings that
start and end on threebranes were constructed. These correspond to BPS states
of the ${\cal N}=4$ super Yang-Mills theory. These authors argue for the same 
cut off employed in our work.

{\it Acknowledgements:} It is a pleasure to thank Antal Jevicki, Jo\~ao Nunes
and Sanjaye Ramgoolam for useful discussions. The work of RdMK is supported by
a South African FRD bursary. The work of APC and JPR is partially 
supported by the FRD under grant number GUN-2034479.

\listrefs
\vfill\eject
\bye